\documentstyle[twocolumn,aps,epsf]{revtex}
\begin{document}
\bibliographystyle{roman}
\def\hb{\hfill\break}
\def\MeV{\rm MeV}
\def\GeV{\rm GeV}
\def\TeV{\rm TeV}

\def\m{\rm m}
\def\cm{\rm cm}
\def\mm{\rm mm}
\def\lam{$\lambda_{\rm int}$}
\def\rad{$X_0$}
 
\def\NIM{Nucl. Instr. and Meth.~}
\def\etal{{\it et al.~}}
\def\eg{{\it e.g.,~}}
\def\ie{{\it i.e.~}}
\def\cf{{\it cf.~}}
\def\etc{{\it etc.~}}
\def\vs{{\it vs.~}}
\begin{minipage} [t]{6.7in}
\Large 
\centerline{\bf On Big Bang Relics, the Neutrino Mass and the Cosmic Ray Spectrum}
\vskip 4mm
\normalsize
\centerline{Richard WIGMANS}
\vskip 2mm
\centerline{\em Department of Physics, Texas Tech University, Lubbock TX 79409-1051, USA}
\vskip 2mm
\centerline{(Submitted to Phys. Rev. Lett. on 13 November 1998)} 
\vskip 3mm

\normalsize

\begin{abstract}
It is shown that high energy features of the cosmic ray spectrum, in
particular the kink around 4 PeV and the corresponding change in spectral index, may be
explained from interactions between highly energetic protons and relic Big Bang
antineutrinos, if the latter have a rest mass of about 0.4 eV/$c^2$. This explanation is 
supported by experimental data from extensive air-shower experiments.\hb\hb
\end{abstract}
\end{minipage}
\vskip 8mm

According to the Big Bang model of the evolving Universe, large numbers of (electron) neutrinos
and antineutrinos have been around since the beginning of time. During the very first second,
these (anti-)neutrinos were in thermal equilibrium with the other particles that made up the
primordial soup: photons, electrons, positrons and nucleons. Photon-photon interactions
created $e^+e^-$ pairs, which annihilated into photon pairs. Interactions between
(anti-)neutrinos and nucleons turned protons into neutrons and vice versa.

This {\em leptonic era} came to an end when this thermal equilibrium ceased to exist.
In the reaction
\begin{equation}
\bar{\nu}_e + p \rightarrow n + e^+
\label{primo}
\end{equation}
the mass difference between the particles in the final and initial states amounts to 1.8 MeV.
For this reaction to occur, the energy available in the $\bar{\nu}_e p$ center-of-mass
system has to exceed this value.
When the temperature dropped below 1 MeV,
reaction (1) became energetically impossible in most $\bar{\nu}_e p$ encounters.
In addition, because of the expansion of the Universe and the corresponding reduced density,
the mean free path of the neutrinos became increasingly long, and
soon the average neutrino lifetime exceeded the age of the Universe.
The thermal equilibrium between neutrinos and other particles thus came to an end, the
neutrinos {\em decoupled} and the neutron/proton ratio froze out, determining the
primordial He/H ratio.

Since that moment, the wavelengths of the neutrinos have been expanding in
proportion to the size of the Universe. Their present spectrum is believed
to be a Fermi--Dirac distribution characterized by a temperature of 1.95 K\cite{TG}.
The present density of these Big Bang relics is estimated at $\sim$ 100 cm$^{-3}$ 
for each neutrino flavor, nine orders of magnitude larger than the density of baryons in the
Universe. 

In spite of this enormous density, relic neutrinos have until now escaped direct
detection. They are acknowledged as a potentially very important component of dark matter, but
this hinges on their mass, which is not known either. 
The recent observation of 
a zenith-angle dependence in the ratio of $\nu_e/\nu_\mu$
induced interactions in the Superkamiokande detector\cite{Superk} is generally considered
strong

\centerline{  }
\vskip 56mm\parindent=0mm 
evidence for (at least one) non-zero neutrino restmass, and suggests a {\em lower} limit
of the order of 0.03 eV.
The {\em upper} limit on the $\nu_e$ mass\footnote{In the following, we express masses in 
energy units, omitting the $c^{-2}$ factor.}
is 15 eV\cite{PDG} and comes from endpoint measurements of tritium
$\beta$ decay and from the measured spread in arrival times of $\nu_e$'s from supernova
SN1987a. 

\parindent=3mm
The reason why relic neutrinos have yet to be observed is their
extremely small kinetic energy, which makes it hard to find a process through which
they might reveal themselves. In this {\em Letter}, such a process is proposed and its
consequences are examined.
\vskip 2mm

Let us imagine a target made of relic $\bar{\nu}_e$'s and let us bombard this target with
protons. 
Let us suppose that we can tune this imagined beam to arbitrarily high energies.
Then, at some point, the proton energy will exceed the value at which the
$p \bar{\nu}_e$ center-of-mass energy exceeds the combined mass
of a neutron and a positron. Beyond that threshold, the inverse $\beta$-decay reaction
\begin{equation}
p + \bar{\nu}_e \rightarrow n + e^+
\label{ibeta}
\end{equation}
is energetically possible.
The threshold proton energy for this process depends on the mass of the $\bar{\nu}_e$
target particles. If this mass is large compared to the kinetic energy of
the target particles, this may be considered a stationary-target problem, and
the center-of-mass energy can be written as
\begin{equation}
E_{\rm cm}~=~\sqrt{m_p^2 + m_\nu^2 + 2 E_p m_\nu}~\approx~\sqrt{m_p^2 + 2 E_p m_\nu} 
\label{com1}
\end{equation}
When the experimental mass value of the proton (938.272 MeV) is substituted in
eq. \ref{com1} and the requirement is made that $E_{\rm cm} > m_n + m_e$ (940.077 MeV), this
leads to  
\begin{equation}
E_p m_\nu > 1.7 \cdot 10^{15}~ {\rm (eV)}^2 
\label{com2}
\end{equation}
Thus, for a neutrino mass of 1 eV, this process may take place for proton energies in excess of
1.7 PeV, while the threshold energy is 17 PeV when the mass is only 0.1 eV.

Obviously, the target needs to be huge to see any effect 
of the described process. The largest imaginable $\bar{\nu}_e$ target
is the Universe. This target is continuously traversed by cosmic rays, 
90\% of which are free protons (the remaining fraction consists predominantly of $\alpha$
particles).
\centerline{\  }
\begin{figure}[htb]
\vspace{-0.3cm}
\hbox{\epsfxsize=86mm 
\epsffile{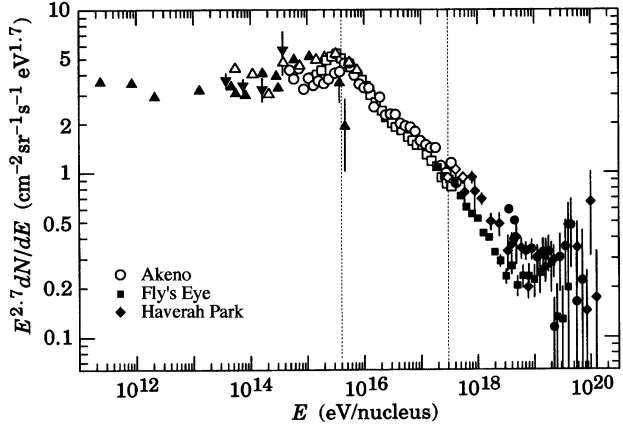}
}
\caption[]{\small
The all-particle spectrum of charged cosmic rays}
\label{Kink1}
\end{figure}
\noindent

Figure \ref{Kink1} shows the spectrum of these cosmic rays, compiled by the
Particle Data Group \cite{PDG2}. This spectrum falls very steeply with energy.
At energies below 1 PeV, the differential energy spectrum is well described by
\begin{equation}
{dN\over dE}~=~1.8~ E^{-2.7}~{\rm nucleons\over{cm^2~sr~~s~~GeV}}
\label{cosray}
\end{equation}
In order to display characteristic features of this spectrum which would
otherwise be hard to discern, the differential energy spectrum has been
multiplied by $E^{2.7}$ in this figure. The steepening that occurs between 1 PeV and 10 PeV
is known as the {\em knee} of the cosmic ray spectrum. Ever since its discovery, this
feature has been  the subject of intense interest and many explanations have been proposed.
Especially models in
which the cosmic rays are the result of particle acceleration in the shock waves produced in
supernova explosions (the Fermi mechanism) have received much attention in the 
literature. Such models predict a maximum energy, proportional to the nuclear
charge $Z$ of the particles\cite{Blanford}.  

It is expected that the cosmic ray spectrum exhibits a high-energy cutoff at $\sim 10^{20}$ 
eV\cite{GZK}. Beyond that threshold, photopion production in collisions between cosmic protons
and microwave background photons is energetically possible. Therefore, one does not not
expect to observe any such protons, unless their source is located relatively nearby 
(\ie , $< 50$ Mpc). 

\vskip 2mm
The inverse $\beta$-decay reaction (2) could, in principle, lead to a similar
depletion of protons at energies above the threshold. If this process were indeed responsible
for the change in the spectral index observed around 4 PeV, the neutrino mass should thus have
a value of about 0.4 eV (4).
We therefore postulate the following hypothesis: 

{\em The mass of the relic electron neutrinos is 0.4
eV. The change of the spectral index in the all-particle cosmic ray spectrum at an energy of 
$\sim 4 \cdot 10^{15}$ eV is caused by the onset of the reaction $p + \bar{\nu}_e \rightarrow
n + e^+$, which becomes energetically possible at this energy.}
\vskip 2mm

Let us examine this hypothesis. The most crucial question concerns the {\em probability} of
the $p \bar{\nu}_e$ reactions. 
In their historic experiment, Cowan and Reines measured the cross section for the inverse
process (1), at energies just above
the threshold, to be $\sim 10^{-43}$ cm$^2$ \cite{Reines}.
As the proton energy increases above the threshold value, the cross section for the reaction
$p + \bar{\nu}_e \rightarrow n + e^+$ rises proportional to the {\em available} energy in the
center-of-mass system,
\ie the energy above the $Q$-value of 1.8 MeV\cite{Perkins}:
\begin{equation}
\sigma_{p\bar{\nu}_e \rightarrow ne^+}~=~0.34 (E_{\rm cm} - Q) 10^{-41} {\rm cm}^2
\label{xsec}
\end{equation}
with the energies expressed in units of MeV. For proton energies between $10^{16}$ eV and
$10^{18}$ eV, the cross section thus rises from $10^{-41} - 10^{-39}$ cm$^{2}$.
At a target density of $10^2$/cm$^3$, this cross section translates for a $10^{16}$ eV
proton in a  mean free path of $10^{39}$ cm, or an average lifetime of
$10^{21}$ years, 11 orders of magnitude longer than the age of the Universe. Therefore, it
seems very unlikely that the proposed process could affect the cosmic proton spectrum at a
significant level. 

\vskip 2mm

However, it is important to realize that, with a mass of 0.4 eV, the relic
Big Bang $\bar{\nu}_e$'s would be {\em nonrelativistic} ($kT \sim 10^{-4}$ eV). 
Typical velocities would be of the order of $10^5$ m/s in that case\cite{TG}, less than the 
escape velocity from the surface of the Sun.
Such neutrinos may be expected to have accumulated in gravitational potential wells.
Weiler\cite{weiler} recently estimated that the density of relic neutrinos in our own galaxy
would increase by four orders of magnitude (compared to the universal density of 100/cm$^3$)
if their mass were 1 eV.

Locally, this effect could be much more spectacular. Extremely dense objects, such as neutron
stars and black holes could accumulate very large numbers of relic neutrinos and
antineutrinos in their gravitational field. 
Let us consider, as an example, a typical neutron star, with a mass ($M$) of $3 \cdot 10^{30}$
kg and a radius of 10 km. Even at a distance ($r$) of one million kilometers from this object,
the escape velocity is still considerably larger than the typical velocity of these relic
neutrinos: 700 km/s.

The concentration of relic neutrinos in such a local potential well is ultimately
limited by the Pauli principle, which limits their phase-space density to $4 g_\nu h^{-3}$,
where
$g_\nu$ denotes the number of helicity states and $h$ Planck's constant\cite{TG}. 
Since the escape velocity scales with $r^{-1/2}$, the maximum neutrino density is proportional
to $r^{-3/2}$, and reaches values of $\cal{O}$$(10^{12}/{\rm cm}^3)$ near the surface of this
neutron star. If the source of the potential well has a larger mass, the
achievable neutrino densities increase proportional to $M^{3/2}$. In the ``neutrino
atmosphere'' surrounding such dense objects, the lifetime of a $10^{16}$ eV proton could thus
be reduced to values comparable to, or even considerably smaller than the age of these objects.
      
The example of a neutron star was chosen since objects of this type may well play a role in
accelerating protons to the very high energies considered here. Neutron stars
usually rotate very fast and exhibit very high magnetic fields (typically $\sim 10^8$ T).
When the magnetic axis does not correspond to the rotation axis, the changing magnetic
fields in the space surrounding the neutron star may give rise to substantial electric fields,
in which charged particles may be accelerated to high energies\cite{KOT91}.
The strong magnetic fields, in turn, will tend to confine the accelerated particles to the
region surrounding the neutron star. When the charged particles are electrons, the acquired
energy is emitted in the form of synchrotron radiation, a characteristic signature of many
pulsars.

However, for protons this type of energy loss only plays a
significant role at energies that are 13 orders of magnitude ($(\gamma_p/\gamma_e)^4$)
larger than for comparable electrons. For example, a 10$^{16}$ eV proton describing a circular
orbit with a radius of 1,000 km loses per orbit about 0.6\% of its energy in the form of
synchrotron radiation. To put this in perspective, that is about a factor of 5 less than
the fractional energy loss experienced by 100 GeV electrons in CERN's LEP ring. 
It is, therefore, not inconceivable that a stationary situation might arise in which protons
and heavier ions, initially attracted by gravitational forces, and accelerated to high
energies in the electric fields mentioned above, end up in a closed orbit around
the neutron star in which energy gained and synchrotron losses balance each other. In that
case, collisions with relic neutrinos accumulated in the gravitational field of the neutron
star might become a realistic possibility.

The sharpness of the knee suggests that, if our hypothesis is correct, most of the protons
with energies above the threshold trapped in the gravitational field undergo this fate. One
may then wonder why the cosmic ray spectrum does not exhibit a ``step'' at the threshold
energy. 
Two factors are responsible for this.
%

First, there are the effects of the detector response function.
Upward fluctuations in the measured signal have a major impact on
the measured spectrum of a very steeply falling distribution such as this one. Especially when
this response function is not Gaussian, these effects may extent over a considerable
energy range.

Second, the neutrons produced in
(\ref{ibeta}) eventually decay into protons which carry, on average, 1/6 of the energy of the
protons that initiated the process, with a broad distribution about this average. Decay
protons with an energy below 4 PeV contribute to a ``bump'' in the spectrum just below this
threshold. Such a bump is indeed experimentally observed, and is known as the {\em kneecap}
(see fig. \ref{Kink1}). As a matter of fact, this bump contributes significantly to the
sharpness of the knee.  
%

As a result of these two effects, the cosmic ray spectrum would be gradually depleted of
protons at energies beyond the 4 PeV threshold.
\vskip 2mm

The cosmic ray spectrum is a rich source of possible tests of our hypothesis.
For example, if the knee is indeed caused by the onset of the inverse $\beta$-decay
reaction, the observed kink should be a feature for protons and protons alone. But there is
more.

Other cosmic ray particles may interact with the relic background as well.  
The reactions in which $\alpha$ particles are dissociated in collisions with relic
$\nu_e$ and $\bar{\nu}_e$   
\begin{eqnarray}
\alpha + \nu_e~ \rightarrow~ 3p + n + e^-\\
\alpha + \bar{\nu}_e~ \rightarrow~ p + 3n + e^+
\end{eqnarray}
have $Q$-values of 27.5 MeV and 30.1 MeV, respectively. The threshold
energies for these reactions are larger than the threshold energy
for reaction (\ref{ibeta}) by factors of 61 and 66, respectively.
  
The cosmic ray spectrum in the range around the knee was measured in detail by the
Akeno extensive air-shower experiment\cite{akeno}. From a fit to their experimental data,
the kink was located at $\log{E_p}~({\rm eV}) = 15.67$.   
If our explanation of the origin of this kink is correct, a kink in the
spectrum of cosmic $\alpha$ particles should thus be expected at an energy that is larger by
a factor of 61-66: $\log{E_\alpha} \approx 17.5$ ($E_\alpha \approx 3\cdot 10^{17}$ eV). 
By contrast, 
supernova shockwave models predict kinks in the proton and helium spectra
separated by only a factor of two ($Z_{\rm He}/Z_{\rm H}$) in energy. Accurate measurements of
the $p/\alpha$ ratio in the energy range $10^{14} - 10^{18}$ eV should thus make it possible to
obtain conclusive evidence either way.

It is remarkable that several extensive air-shower experiments have reported a kink 
around $\log{E} = 17.5$. The Fly's Eye detector, which obtained the largest event
statistics, observed a change in the spectral index from $3.01 \pm 0.06$ for 
$E < 10^{17.6}$ eV to $3.27 \pm 0.02$ for energies
\begin{figure}[htb]
\vspace{-0.4cm}
\hbox{\epsfxsize=86mm 
\epsffile{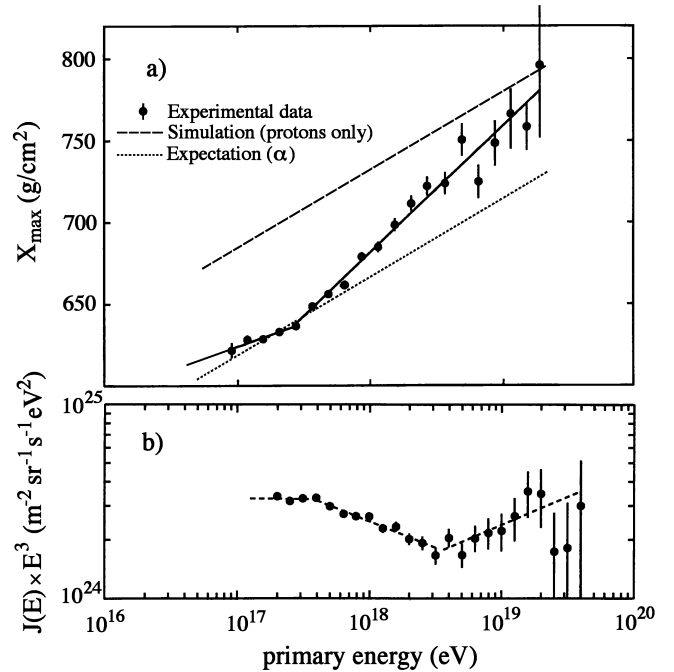}
}
\caption{\small
Experimental results obtained with the Fly's Eye detector\protect\cite{fleye}. See text for
details.}
\label{fly}
\end{figure}
\parindent=0mm
$10^{17.6} < E < 10^{18.5}$ eV\cite{fleye}.
The Haverah Park detector also reported a kink at $\log{E} = 17.6$, with the spectral index
changing from $3.01 \pm 0.02$ to $3.24 \pm 0.07$\cite{haverah}. 

\parindent=3mm
The cosmic ray spectrum obtained with the Fly's Eye detector is shown in fig.
\ref{fly}$b$. 
This detector is also capable of measuring the {\em altitude} (and thus the amount of air
traversed by the shower particles, $X_{\rm max}$) at which the shower maximum is located  (fig.
\ref{fly}$a$). This distribution
exhibits an unmistakable kink at approximately the same energy as the kink in the cosmic ray
spectrum, indicating an abrupt change in chemical composition.
The dashed line in fig.
\ref{fly}$a$ represents the results of a simulation in which the authors modeled the showers
initiated by protons\cite{fleye}. To estimate what the  equivalence of the dashed line will
look like for showers induced by $\alpha$ particles, two effects have to be considered. 
 
First, the interaction length ($\lambda_{\rm int}$) of $\alpha$'s is shorter than
for protons. For protons in air, $\lambda_{\rm int}$ amounts to 90 g/cm$^2$.
Since the total cross section for interactions induced by highly energetic ions with atomic
number
$A$ scales with $A^{2/3}$, $\lambda_{\rm int}$ scales with $A^{-2/3}$, which gives
$\lambda_{\rm int} = 55$ g/cm$^2$ for $\alpha$'s in air, a difference of 35 g/cm$^2$. 

Second, the energy carried by the incoming particle is distributed to a
larger number of secondaries. This has the effect that, in the subsequent shower
development, the shower maximum is reached earlier than for proton showers 
of the same energy. The effect of this can, in first approximation, be estimated by
considering the ion-induced shower as a collection of $A$ showers, each with an energy
$E_A/A$. From the slope of the dashed line in fig. \ref{fly}$a$, one may estimate 
$\alpha$ induced showers in air to reach their maximum earlier than proton induced ones, by 30
g/cm$^2$, counting from the first nuclear interaction.

These (somewhat oversimplified) considerations thus lead to $X_{\rm max}$ values for $\alpha$
induced reactions in the atmosphere that are 65 g/cm$^2$ smaller than those for protons
of the same energies. This result is represented by the dotted curve in fig.
\ref{fly}$a$.    

If the hypothesis described in this {\em Letter} were correct, one would expect the $X_{\rm
max}$ distribution to exhibit the following features.
At energies below $10^{15}$ eV, the $X_{\rm max}$ value follows a line parallel to the ones
for protons and $\alpha$'s, somewhere in between these two, depending on the relative H/He
abundances.
At $4 \cdot 10^{15}$ eV, the threshold for reaction (2) is crossed. Beyond that
point, protons gradually disappear from the spectrum. The $X_{\rm max}$ curve thus shows a
kink at this point, beyond which the slope is
less steep than before. For energies around
$10^{17}$ eV, almost all protons have disappeared and the spectrum is dominated by
$\alpha$ particles. At $\sim 3 \cdot 10^{17}$ eV, a new threshold is crossed,
this time for reactions (7) and (8). Beyond that point, $\alpha$'s gradually disappear from the
spectrum. In that process, protons carrying a significant fraction of the energy are produced.
This leads to a gradual enrichment in protons (\ie, to a slope in the
$X_{\rm max}$ distribution that is steeper than that for protons or $\alpha$'s) . Beyond
$10^{18.5}$ eV, protons of extragalactic origin add to the mix.

The Fly's Eye's experimental data (fig. \ref{fly}$a$) are in detailed
agreement with the described scenario. Further measurements of $X_{\rm max}$, at energies in
the range $10^{15} - 10^{17}$ eV, would constitute another important test.
 
\vskip 2mm
If the hypothesis decribed in this {\em Letter} turned out to be correct, a new era in cosmic
ray physics may be envisaged. Studies of the multi-PeV proton component of the spectrum would
give a direct window on a very early epoch in the history of the Universe, \ie the
leptonic era. 
Proof of the existence of relic neutrinos would be a very strong additional argument
in favor of the Big Bang model. Further measurements of their properties could provide more
detailed insight in the processes that shaped the Universe in its first second.
A neutrino mass of 0.4 eV would crucially increase the understanding of dark-matter issues. 
A 0.4 eV neutrino mass would also provide new perspectives for neutrino astronomy. For
example, if a supernova occurred at a distance of 1 Mpc, the arrival times of simultaneously
produced neutrinos of 1 MeV and 10 MeV would differ by about 7 seconds. Neutrino astronomy
might one day even provide a precision tool for distance measurements on this basis.

The high-energy cosmic proton and $\alpha$ spectra may thus hold the key to an important
breakthrough in our understanding of the physical world.

\bibliographystyle{unsrt}

\end{document}